\pdfoutput=1

\documentclass[prl,twocolumn,showpacs,preprintnumbers,amsmath,amssymb,
superscriptaddress,nofootinbib]{revtex4-1}
\usepackage{epsfig}
\usepackage{graphicx}
\usepackage{color}

\newcommand{\smallminus}{{\rm\rule[2.4pt]{6pt}{0.65pt}}}
\newcommand{\smallplus}{\hspace{0.5pt}\text{{\small+}}\hspace{-0.5pt}}
\newcommand{\mi}{\smallminus}

\newcommand{\sdots}{\scalebox{0.9}{$\dots$}}

\begin{document}

\preprint{}

\title{New Soft Theorems for Goldstone Boson Amplitudes}

\author{Karol Kampf}
\affiliation{Institute of Particle and Nuclear Physics, Charles University, Prague, Czech Republic}

\author{Jiri Novotny}
\affiliation{Institute of Particle and Nuclear Physics, Charles University, Prague, Czech Republic}

\author{Mikhail Shifman}
\affiliation{William I. Fine Theoretical Physics Institute, University of Minnesota, Minneapolis, MN, USA}

\author{Jaroslav Trnka}
\affiliation{Center for Quantum Mathematics and Physics (QMAP), University of California, Davis, CA, USA}
\affiliation{Institute of Particle and Nuclear Physics, Charles University, Prague, Czech Republic}


\begin{abstract}
In this letter we discuss new soft theorems for the Goldstone boson amplitudes with non-vanishing soft limits. The standard argument is that the non-linearly realized shift symmetry leads to the vanishing of scattering amplitudes in the soft limit, known as the Alder zero. This statement involves certain assumptions of the absence of cubic vertices and the absence of linear terms in the transformations of fields. For theories which fail to satisfy these conditions, we derive a new soft theorem which involves certain linear combinations of lower point amplitudes, generalizing the Adler zero statement. We provide an explicit example of $SU(N)/SU(N-1)$ sigma model which was also recently studied in the context of $U(1)$ fibrated models. The soft theorem can be then used as an input into the modified soft recursion relations for the reconstruction of all tree-level amplitudes. \end{abstract}

\maketitle

\vspace{-0.2cm}

\section{Introduction}

\vspace{-0.25cm}

In this paper we connect two different topics which have been intensively studied in last few years: soft limits of scattering amplitudes in effective field theories, and the $U(1)$ fibrated CP$(N\mi1)$ sigma models. The tree-level S-matrix in these models exhibit a very special behavior in the soft limit which gives rise to the new type of soft theorems, distinct from the usual Adler zero.

\smallskip

\noindent {\bf Sigma models}: The $U(1)$ fibrated CP$(N\mi1)$ models represent a
class of sigma models interpolating between CP$(N\mi1)$ and  $S^{2N\mi1}$ target spaces \cite{tu1,tu2,tu3}. These models correspond to the cosets $\left[ \left( SU(N)/SU(N\mi1)\times U(1)\right) \right] \times
U(1) $. For brevity in the following we refer to these models as $SU(N)/SU(N\mi1)$. The above class contains an extremely interesting example of $N=2$, including CP(1) and $S^{3}$ models, both being integrable and exactly
solvable in two spacetime dimensions \cite{tu3,tu4,tu5,W2}. The algebraic form of the interpolating
Lagrangian is 
\begin{equation}
\mathcal{L}=\frac{1}{2\lambda ^{2}} \Bigl\{ \Bigr[ \sum_{a=1,2,3} J_{\mu
}^{a}J^{a \mu} \Bigr] -\kappa J_{\mu }^{3}J^{3 \mu }\Bigr\}\,,  \label{cl1}
\end{equation}
where the current $J_{\mu }$ is defined as 
\begin{equation}
J_{\mu }=U^{\dagger }\partial _{\mu }U\equiv 2\mathrm{i}\,\sum  J_{\mu
}^{a}T^{a}\,,
\quad J_{\mu }^{a}=-\mathrm{i}\,\mathrm{Tr}\,(J_{\mu }T^{a})\,.
\label{cl2}
\end{equation}
Here $U$ is an arbitrary $x$-dependent matrix, $U(x)\in SU(2)$, the
generators are proportional to the Pauli matrices, $T^{a}=\frac 12 \tau ^{a}$, and $\kappa $ is a numerical deformation parameter. If $\kappa =1$ the theory is equivalent to the CP(1) model, while at $\kappa =0$ it reduces to
the  $SU(2)\times SU(2)/SU(2)$ Principal Chiral Model (PCM) whose target space is $S^{3}$. For arbitrary $N$ we can extend (\ref{cl1}) as follows: 
\begin{equation}
\mathcal{L}=\frac{1}{2\lambda ^{2}}
\sum_{a=1}^{2N-2}\Bigl[(J^{N^{2}-2N+a})^{2}
+\frac{1-\kappa}{N}(J^{N^{2}-1})^{2}\Bigr]\,.  \label{general Lagrangian}
\end{equation}
\noindent {\bf Scattering amplitudes}: Recently, there has been a huge progress in new methods for the calculation of on-shell scattering amplitudes in QFTs. While most work has
been focused on gauge theory and gravity, especially with maximal
supersymmetry, new surprising results have been obtained in the case of
effective field theories (EFT). The general approach is to fix the amplitude uniquely by imposing certain sets of constraints. The universal example is a tree-level factorization on poles,
\begin{equation}
\lim_{P^{2}\rightarrow 0}A_{n}=\sum\frac{A_{L}A_{R}}{P^{2}}\,,
\label{fact}
\end{equation}
where the sum runs over internal states. The set of all factorizations is enough to completely specify tree-level S-matrix in a large class of QFTs, called {\it on-shell constructible}, including gauge theories or gravity, and it can be then calculated
using the recursion relations \cite{recur}. 

This does not apply to EFTs due to the presence of unfixed contact terms with no
poles, which originate from higher-dimensional operators in the Lagrangian.
In \cite{Cheung:2015ota} it was shown that when the amplitude vanishes 
for one of the momenta going to zero, we can impose this
information as a constraint and use \emph{soft recursion relations} for
on-shell reconstruction. This singles out a set of exceptional EFTs where
all coefficients in the Lagrangian are fixed by the requirement of a certain
degree of vanishing, $A_{n}=\mathcal{O}(p^{\sigma })$, in the soft limit 
\cite{Cheung:2014dqa,Cheung:2016drk,Cheung:2018oki,Elvang:2018dco,Cheung:2015ota,Low:2019ynd,unique}.

The primary example is the PCM describing the spontaneous symmetry breaking $SU(N) \times
SU(N) \rightarrow SU(N)$. It has been known since
1970s \cite{Susskind:1970gf} that the requirement of the vanishing soft
limit of amplitudes, known also as the \emph{Adler zero}, on any
two-derivate theory specifies NLSM as a unique solution. In \cite{NLSM} 
it was found that in this model the group part of tree-level amplitudes can be stripped, similar to Yang-Mills amplitudes, dramatically simplifying  the calculations.

\vspace{-0.25cm}

\section{Adler zero}

\vspace{-0.25cm}

First we review the standard textbook derivation of the Adler zero for
amplitudes of Nambu-Goldstone bosons (NGB). We start with the theory for the
single NGB corresponding to the spontaneous breaking of one-parameter
continuous symmetry. The NGB couples to the associated Noether current $N^{\mu }(x)$ 
with a strength parametrized by the decay constant $F$, 
\begin{equation}
\langle 0|N^{\mu }(x)|\phi (p)\rangle =-\mathrm{i}p^{\mu }Fe^{-\mathrm{i}
p\cdot x} \,.
 \label{mat1}
\end{equation}
The matrix element of this current between physical states has a pole for 
$p^{2}\rightarrow 0$, and the residue corresponds to the scattering amplitude
for the NGB emission. For the element between out state 
$\langle \alpha |$ and in state $|\beta \rangle $ we get 
\begin{equation}
\langle \alpha |N^{\mu }(0)|\beta \rangle =F\frac{p^{\mu }}{p^{2}}
A_{n}(\alpha +\phi (p),\beta )+R^{\mu }(p).
\label{mat3}
\end{equation}
Here $A_{n}(\alpha +\phi (p),\beta )$ is the on-shell amplitude which
involves emission of the state $\phi $ with momentum $p$, where $p^{\mu
}=P_{\beta }^{\mu }-P_{\alpha }^{\mu }$ is the difference between incoming
and outgoing momenta, and $R^{\mu }(p)$ is the regular function for 
$p^{2}\rightarrow 0$. Due to the conservation of the current we have 
$p_{\mu }\langle \alpha |N^{\mu }(0)|\beta \rangle =0$ and therefore,
\begin{equation}
A_{n}(\alpha +\phi (p),\beta )=-\frac{1}{F}p_{\mu }R^{\mu }(p)\,.
\label{mat4}
\end{equation}
Suppose that $R^{\mu }(p)$ is regular also in the limit $p\rightarrow 0$. 
This is an additional assumptions which does not follow
automatically from the standard polology. Then the amplitude $A_{n}$
vanishes if the NGB momentum is soft, 
\begin{equation}
\lim_{p\rightarrow 0}A_{n}(\alpha +\phi (p),\beta )=0\,. 
\end{equation}
This is the statement of the Adler zero. The same argument applies to the
theory with multiple Goldstone bosons. To summarize, we have the
nonperturbative Adler zero provided the matrix element $\langle \alpha
|N^{\mu }(0)|\beta \rangle $ of the Noether current corresponding to the
spontaneous symmetry breaking \emph{has no other singularity for\/
$p\rightarrow 0$ besides the NGB pole\/}. Therefore, the violation 
of the Adler zero is possible only when there are
additional singularities in the matrix element of the Noether current. This
is achieved in the case when the Noether current can be inserted into the
external lines of the amplitude $A_{n}(\alpha ,\beta )$, i.e. when there are
quadratic terms in the expansion of the operator $N^{\mu }$ in the
elementary fields. There are two sources of these quadratic terms:

\vspace{-1mm}

\begin{enumerate}
\item The presence of cubic vertices in the Lagrangian;

\vspace{-1mm}

\item The presence of linear terms in the nonlinearly realized symmetry
transformation corresponding to the Noether current $N^{\mu }$.
Schematically, 
$$
\delta \phi =a+b\phi +\mathcal{O}(\phi ^{2}),\qquad b\neq 0.
$$
\end{enumerate}

\vspace{-0.2cm}

These two conditions are \emph{not} sufficient: even when at least one of
the above conditions is satisfied, the theory can still have the Adler zero --
a more detailed analysis is needed. Note that the cubic vertices can be
always removed by means of field redefinitions, as there are no on-shell
three-point amplitudes (apart from $\phi ^{3}$ theory). In such a case the
presence of the linear term in $\delta \phi $ is crucial. Note that e.g. in
the PCM parametrized by the Lagrangian 
\begin{equation}
\mathcal{L}=F^{2}\mathrm{Tr}(\partial ^{\mu }U^{\dagger })(\partial _{\mu
}U),\quad U=e^{\frac{\mathrm{i}}{F}\phi },\;\phi =\phi ^{a}T^{a}\,,
\label{NLSM}
\end{equation}
where $U\in SU(N)$ transforms under the general element $(V_{R},V_{L})$ of
the chiral group $SU(N)\times SU(N)$ as 
\begin{equation}
U\rightarrow V_{R}UV_{L}^{-1}\,,  \label{chiral}
\end{equation}
there are no cubic vertices, and the matrix $\phi$ of $N^{2}-1$ scalar
fields transforms under the axial transformation 
$V_{L}=V_{R}^{-1}=1+\mathrm{i}\alpha ^{a}T^{a}\equiv 1+\mathrm{i}\alpha $ (with $\alpha$ infinitesimal) as  
\begin{equation}
\delta _{\alpha }\phi =2F\alpha -\frac{1}{6F}\{\alpha ,\phi ^{2}\}+\frac{1}{3F}\phi \alpha \phi +O\left( \phi ^{3},\alpha ^{2}\right) .
\end{equation}
The linear term is absent, and consequently the theory has the Adler zero. 

\vspace{-0.25cm}

\section{New soft theorem}

\vspace{-0.2cm}

Let us assume a general two-derivative Lagrangian for $N$ fields $\left\{
\phi _{I}\right\} _{I=1}^{N}$ with a cubic vertex, 
\begin{equation}
\mathcal{L}=\frac{1}{2}\partial_\mu \phi _{I}\partial^\mu \phi _{I}+
\frac{1}{2}K_{IJK} \partial_\mu \phi _{I} \partial^\mu \phi _{J}\phi _{K}+
\mathcal{O}(\phi ^{4})  \label{Lagr2}
\end{equation}
with sum over repeating indices tacitly assumed. Let the transformation of the fields
corresponding to spontaneously broken symmetry contain, besides the constant
term, also a linear term, 
\begin{equation}
\delta ^{J}\phi _{I}=F_{I}^{J}+\sum_{K=1}^{N}C_{IK}^{J}\phi _{K}+\mathcal{O}%
(\phi ^{2}).  \label{inv}
\end{equation}
The invariance of (\ref{Lagr2}) under the symmetry (\ref{inv}) requires
non-trivial constraints between all coefficients, namely,
\begin{equation}
B_{IK}^{J}\equiv C_{IK}^{J}+\frac{1}{2}\sum_{L=1}^{N}K_{IKL}F_{L}^{J}
\end{equation}
must be antisymmetric, $B_{IK}^{J}=-B_{KI}^{J}$. The Noether current 
$N_{\mu }^{J}$ contains a quadratic term in the field expansion 
\begin{equation}
N_{\mu }^{J}=\sum_{I=1}^{N}F_{I}^{J}\partial _{\mu }\phi
_{I}+\sum_{L,K=1}^{N}\mathcal{K}_{LK}^{J}\phi _{K}\partial _{\mu }\phi _{L}+
\mathcal{O}(\phi ^{3})\,,
\end{equation}
where $\mathcal{K}_{IK}^{J}$ depend on both parameters $C$ and $K$
\begin{equation}
\mathcal{K}_{IK}^{J}=C_{IK}^{J}+\sum_{M=1}^{N}F_{M}^{J}K_{MIK}.
\end{equation}
At the tree-level the matrix element $\langle \alpha |N_{\mu }^{J}|\beta
\rangle $ has additional singular terms from inserting the
current into external legs. The remainder $R_{\mu }^{J}$ is not regular for 
$p\rightarrow 0$, hence the soft limit of $p^{\mu }R_{\mu }^{J}$ is 
non-zero and reduces to 
\begin{equation}
\lim_{p\rightarrow 0}p^{\mu }R_{\mu }^{J}=-\sum_{L\in \alpha \cup \beta
}\sum_{K=1}^{N}\mathcal{C}_{LK}^{J}A_{n-1}^{K,L}\left( \alpha ,\beta \right),
\end{equation}
where the $A_{n-1}^{K,L}(\alpha ,\beta )$ is the $(n\mi1)$pt amplitude,
the particle $\phi _{L}\left( p_{L}\right) $ is omitted and is replaced by
particle $\phi _{K}\left( p_{L}\right) $ with momentum $p_{L}$. The sum over 
$L$ is over the indices of all the particles in the in and out states.
Therefore the soft theorem has the form 
\begin{equation}
\lim_{p\rightarrow 0}\sum_{I=1}^{N}F_{I}^{J}A_{n}(\alpha +\phi _{I}(p),\beta
)=\sum_{I\in \alpha \cup \beta }\sum_{K=1}^{N}\mathcal{C}_{IK}^{J}A_{n-1}^{K,I}\left( \alpha ,\beta \right) .  \label{soft}
\end{equation}
Here the coefficient function $\mathcal{C}_{IK}^{J}$ is related to the
original parameters in the Lagrangian and transformation as 
\begin{equation}
\mathcal{C}_{IK}^{J}=B_{IK}^{J}+\frac{1}{2}
\sum_{M=1}^{N}F_{M}^{J}(K_{MIK}-K_{MKI})=-\mathcal{C}_{KI}^{J}.
\label{coef2}
\end{equation}
However, since the on-shell amplitudes are invariant with respect to
redefinition of the fields of the form $\phi _{I}=\phi _{I}^{\prime
}+O\left( \phi^{\prime\,2}\right) $, the constants $\mathcal{C}_{IK}^{J}$ do not
depend on such a reparametrization of the Lagrangian. Note that several
conditions must be satisfied in order to get non-zero right hand side of~(\ref{soft}):

\vspace{-0.1cm}

\begin{enumerate}

\item  The coefficients $\mathcal{C}_{IK}^{J}$ must be
non-zero, i.e. no cancellation between parameters in (\ref{Lagr2}), (\ref{inv}) occurs.

\vspace{-0.1cm}

\item The theory needs to have both even and odd amplitudes, as the
amplitudes on the right hand side have $(n\mathrm{\rule[2.4pt]{6pt}{0.65pt}}%
1)$ external legs. Most sigma models do have only even point amplitudes and
therefore, they preserve the Adler zero.

\end{enumerate}

\section{Example of the sigma model}

As an explicit example we consider a theory of two types of NGB fields: a
vector of multiple complex scalar fields $\Phi _{I}^{+}$, $I=1,\dots,N\mi1$, and a single real
scalar $\chi $. We use the parametrization 
\begin{equation}
\hat{u}=\left( 
\begin{array}{c}
\frac{\Phi ^{+}}{F} \\ 
\sqrt{1-\frac{\Phi ^{-}\cdot \Phi ^{+}}{F^{2}}}
\end{array}\right) \,,
\end{equation}
where $\Phi ^{+}=(\phi _{1}^{+},\phi _{2}^{+},\dots ,\phi _{N-1}^{+})^{T}$, $
\Phi ^{-}=\left[ \Phi ^{+}\right] ^{\dagger }$ and $\cdot $ stands for the
contraction over the $I$ index. The Lagrangian of the model is 
\begin{align}
\mathcal{L}& =\frac{(\partial \chi )^{2}}{2}\smallplus\,F^{2}(\partial ^{\mu }\hat{u}^{\dagger }\!\cdot\! \partial
_{\mu }\hat{u})\smallplus\,
\frac{\mathrm{i}F_{0}}{2}\,\partial ^{\mu }\chi (\partial _{\mu }\hat{u}^{\dagger
}\cdot u\,\smallminus\,\hat{u}^{\dagger }\cdot
\partial _{\mu }\hat{u})  \nonumber \\
& -\Bigl( F^{2}\,\mathrm{\rule[2.4pt]{6pt}{0.65pt}}\,\frac{F_{0}^{2}}{2}
\Bigr) (\hat{u}^{\dagger }\cdot \partial ^{\mu }\hat{u})(\partial _{\mu }
\hat{u}^{\dagger }\cdot \hat{u}).  \label{Lagr1}
\end{align}
It has two coupling constants $F$, $F_{0}$ which play the role
of the decay constants of the NGB $\phi _{I}^{+}$ and $\chi $ respectively.
The model described by (\ref{Lagr1}) is a different parametrization of the $SU(N)/SU(N-1)$ non-linear sigma
model (\ref{general Lagrangian}). The relation with the original
couplings is 
\begin{equation}
F_{0}=\frac{1}{\lambda }\left( 1-\kappa \right) ^{1/2},~~~~F=\frac{1}{\sqrt{2%
}\lambda }.
\end{equation}
Let us briefly summarize limiting cases of our model (for details and
discussion see \cite{future}). The limit $\kappa \rightarrow 1$ gives $
F_{0}\rightarrow 0$ and $\chi $ decouples: we get CP$(N\mi1)$ model. The case $
\lambda \rightarrow 0$ with $1-\kappa =O(\lambda ^{2})$ means $F\rightarrow
\infty $ , $F_{0}$ finite and the theory is free. The limit $\kappa
\rightarrow 0$, $\lambda $ fixed means $F_{0}=\sqrt{2}F$ which gives $
O(2N)/O(2N\mi1)$ model. 

Note that the model (\ref{Lagr1}) satisfies the first condition for the Adler zero
violation as it involves the cubic term 
\begin{equation}
\mathcal{L}\ni \mathrm{i}\frac{F_{0}}{2F^{2}}\partial ^{\mu }\chi
\,(\partial _{\mu }\phi _{I}^{-}\cdot \phi _{I}^{+}-\partial _{\mu }\phi
_{I}^{-}\cdot \phi _{I}^{+})\,.
\end{equation}
The Lagrangian is derivatively coupled in the $\chi $ field, and it is
therefore trivially invariant under the shift symmetry 
\begin{equation}\label{eqshiftchi}
\delta \chi = a \,. 
\end{equation}
Since the cubic vertices can be eliminated by the re\-parametrization $\Phi
^{\pm }=\Phi ^{\pm \prime }\exp \bigl( \pm i\frac{F_{0}}{2F}\chi \bigr)$, which does not spoil this property, all scattering amplitudes have
the vanishing soft limit at $p_{\chi }\rightarrow 0$, i.e. for $\chi $ the
Adler zero is valid. 
After this reparametrization, the Lagrangian is also invariant under a more
complicated transformation involving the linear terms, 
\begin{align}
& \delta \chi =\frac{F_{0}}{2F^{2}}\left( a_{I}^{-}\cdot \phi
_{I}^{+}+a_{I}^{+}\cdot \phi _{I}^{-}\right) +\mathcal{O}((\chi ,\phi ^{\pm
})^{2})\,,  \nonumber \\
& \delta \phi _{I}^{\pm }=\mp \mathrm{i}a_{I}^{\pm }\Bigl( 1\mp \frac{F_{0}
} {2F^{2}}\chi \Bigr) +\mathcal{O}((\chi ,\phi ^{\pm })^{2})\,,  \label{sym}
\end{align}
where we introduced shift parameters $a_{I}^{\pm }$. Note that the symmetry
mixes the single scalar field $\chi$ and multiple scalars $\phi _{I}^{\pm }$.
Calculating $\mathcal{C}_{IK}^{J}$ in (\ref{coef2}) we learn that $\mathcal{C}_{IK}^{J}$ is
non-zero. Furthermore the model involves both odd and even amplitudes, and
therefore, the scattering amplitude does not vanish when the momentum of one of 
 $\phi _{I}^{\pm }$ is taken soft. Because of the form of the Lagrangian 
 (\ref{Lagr1}) the only allowed amplitudes have the same number of $\phi ^{+}$
and $\phi ^{-}$ fields. If we think about $\phi ^{\pm }$ as charged scalars,
this just stands for charge conservation. Let us consider now 
the scattering amplitude of $2n$ fields $\phi _{I}^{\pm }$ and $m$ fields $\chi$, 
with total $M=2n+m$ external legs, 
$$
\includegraphics[scale=0.38]{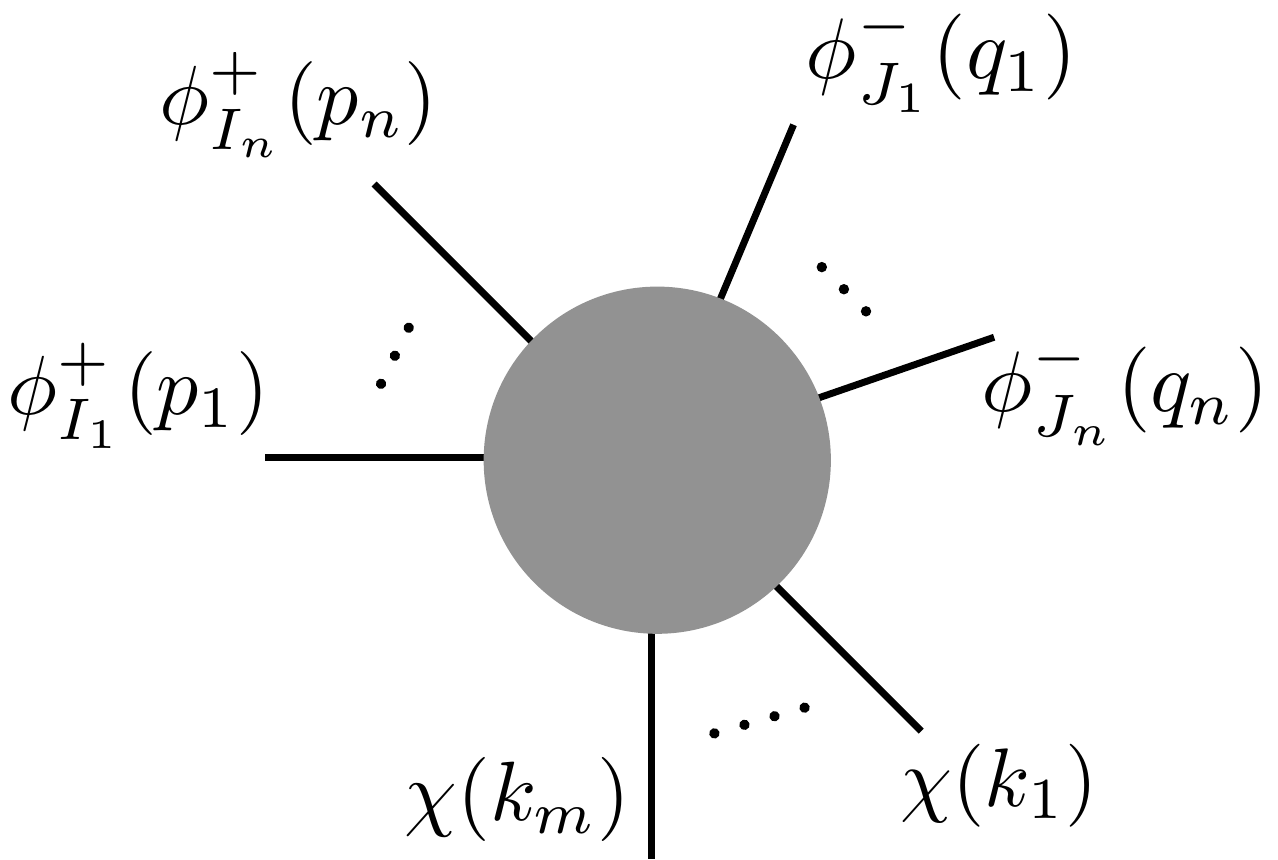} 
$$
\begin{align}
& A_{M}(\{\phi _{I_{i}}^{+}\},\{\phi _{J_{j}}^{-}\},\{\chi \})\equiv
\label{ampl} \\
& A(\phi _{I_{1}}^{+}(p_{1})\sdots \phi _{I_{n}}^{+}(p_{n}),\phi
_{J_{1}}^{-}(q_{1})\sdots \phi _{J_{n}}^{-}(q_{n}),\chi (k_{1})\sdots \chi
(k_{m})).  \nonumber
\end{align}
The soft theorem when the $p_{1}\rightarrow 0$ then reads 
\begin{equation}
\lim_{p_{1}\rightarrow 0}A_{M}=\frac{\mathrm{i}F_{0}}{2F^{2}}
\sum_{i=1}^{m}A_{M-1}^{(i)}-\frac{\mathrm{i}F_{0}}{2F^{2}}
\sum_{j=1}^{n}\delta _{I_{1}J_{j}}A_{M-1}^{(j)}\,,  \label{soft2}
\end{equation}
where the lower point amplitudes are defined as follows: 
\begin{align}
& A_{M-1}^{(i)}\equiv A(\phi _{I_{1}}^{+}(k_{i})\sdots \phi
_{I_{n}}^{+},\{\phi _{J_{j}}^{-}\},\chi (k_{1})\sdots \widehat{\chi (k_{i})}
\sdots \chi (k_{m}))  \nonumber \\
& A_{M-1}^{(j)}\equiv A(\phi _{I_{2}}^{+}\sdots \phi _{I_{n}}^{+},\phi
_{J_{1}}^{-}\sdots \widehat{\phi _{J_{j}}^{-}(q_{j})}\sdots \phi
_{J_{n}}^{-},\chi (q_{j}),\{\chi \}) . \nonumber
\end{align}
In the first case, $A_{M-1}^{(i)}$, we start with $A_{M}$ defined in 
(\ref{ampl}) and remove particle $\chi (k_{i})$, then we replace the particle 
$\phi _{I_{1}}^{+}(p_{1})$ by $\phi _{I_{1}}^{+}(k_{i})$, i.e. just replace
momenta keeping the quantum numbers the same, and finally sum over all
particles $\chi (k_{i})$ which are removed. In the case of $A_{M-1}^{(j)}$
we remove particle $\phi _{I_{1}}^{+}$ completely as well as $\phi
_{J_{j}}^{-}$, and add a new single scalar particle $\chi (q_{j})$ with the
momentum of removed $\phi ^{-}$ particle. Graphically we have (left picture
corresponds to $A^{(i)}$ while the right for $A^{(j)}$) 
$$
\includegraphics[scale=0.32]{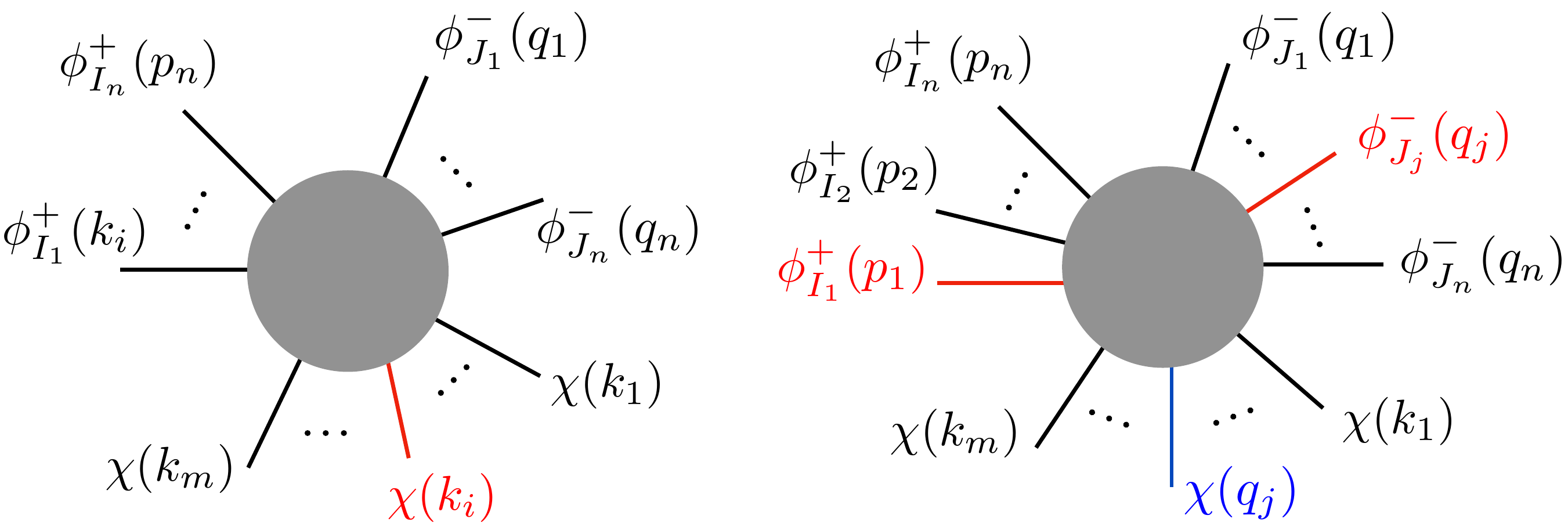} 
$$
where the red color stands for removed legs and blue for the
legs added. For $q_{1}\rightarrow 0$ the soft theorem is the same except the
overall sign on the right hand side of (\ref{soft2}). As discussed earlier any
amplitude vanishes for $k_{j}\rightarrow 0$.

In the following, we focus now on the $N=2$ case which describes only three
fields: $\phi ^{\pm }$, $\chi $. To check the soft theorem we first
calculate all non-vanishing 4pt amplitudes, 
\begin{align}
& A_{4}(\phi _{1}^{+},\phi _{2}^{+},\phi _{3}^{-},\phi _{4}^{-})=
\frac{1}{4F^{4}}(3F_{0}^{2}-8F^{2})s_{12}\,,  \nonumber \\
& A_{4}(\phi _{1}^{+},\phi _{2}^{-},\chi _{3},\chi _{4})=\frac{F_{0}^{2}}{4F^{4}}s_{12}\,,
\end{align}
where $s_{ij}=(p_i+p_j)^2$ and we used the notation $\phi _{1}^{+}\equiv \phi ^{+}(p_{1})$ etc., for simplicity. There is only one non-trivial 5pt amplitude, 
\begin{equation}
 A_{5}(\phi _{1}^{+},\phi _{2}^{+},\phi _{3}^{-},\phi
_{4}^{-},\chi _{5})=\frac{\mathrm{i}F_{0}}{F^{6}}\Bigl( F^{2}\,\mi\,\frac{F_{0}^{2}}{2}\Bigr)(s_{12}\mi s_{34}).\! \label{amp5}
\end{equation}
The soft theorem (\ref{soft2}) for $p_{1}\rightarrow 0$ predicts, 
\begin{align}
\lim_{p_{1}\rightarrow 0}A_{5}& =\frac{\mathrm{i}F_{0}}{2F^{2}}A_{4}(\phi
_{5}^{+},\phi _{2}^{+},\phi _{3}^{-},\phi _{4}^{-})  \label{softT} \nonumber \\
& -\frac{\mathrm{i}F_{0}}{2F^{2}}\left[ A_{4}(\phi _{2}^{+},\phi
_{3}^{-},\chi _{4},\chi _{5})\smallplus\,A_{4}(\phi _{2}^{+},\phi _{4}^{-},\chi
_{3},\chi _{5})\right]  \nonumber \\
&=-\frac{\mathrm{i}F_{0}}{F^{6}}\Bigl( F^{2}\,\mi\,\frac{F_{0}^{2}}{2}\Bigr)s_{34}\,,
\end{align}
in agreement with the direct calculation (\ref{amp5}).

\section{Amplitude reconstruction}

The knowledge of the soft theorem (\ref{soft2}) can be used as an input in the 
modified version of the soft recursion relations introduced in \cite{Cheung:2015ota}. We start with
the momentum shift where all but two particles are shifted in the way that
allows to access the soft limit, 
\begin{align}
& \hat{p_{i}}=p_{i}(1-a_{i}z)p_{i},\quad i=1,\dots n\mi2,  \label{shift1} \\
& \hat{p_{j}}=p_{j}+zq_{j},\quad \quad \,\,\,\,\,j=n\mi1,n, \label{shift2}
\end{align}
where the parameters $a_{i}$ and vectors $q_{j}$ must preserve on-shell conditions
and momentum conservation. For this shift any scattering amplitude scales like 
$A_{n}(z)=\mathcal{O}(z^{2})$, just based on the momentum counting. Then we
consider a residue theorem for the meromorphic function $F_{n}(z)$, 
\begin{equation}
F_{n}(z)\equiv \frac{A_{n}(z)}{z\prod_{i}(1-a_{i}z)}\,. \label{GRT}
\end{equation}
We need at least three factors of $(1-a_{i}z)$ in the denominator to have
vanishing residue at $z\rightarrow \infty$, i.e.
\begin{equation}
\lim_{R\to\infty}\oint_{|z|=R} dz\,F_n(z) = 0 \,.
\end{equation}
We can then express the residue at $z=0$, the original amplitude $A_{n}$, as the sum of all other residues 
\begin{equation}
A_{n}=-\sum_{k}\mathrm{Res}_{z=z_{k}}\,F_n(z)-\sum_{i} \mathrm{Res}_{z=\frac{1}{a_{i}}}F_n(z)\,.
\end{equation}
The first sum on the right hand side refers to factorization poles from 
$A_{n}(z)$, each term is equal to the product of corresponding lower
point amplitudes. The second sum is over the soft limit poles when one of
the $\hat{p_{j}}\rightarrow 0$. In \cite{Cheung:2015ota} we considered only
theories with vanishing soft limits, i.e. the second sum never contributed, but
now the contribution is non-zero and it is given by (\ref{soft}).

As an example, we will reconstruct the 5pt amplitude from $N=2$ model, 
$A_{5}(\phi _{1}^{+},\phi _{2}^{+},\phi _{3}^{-},\phi _{4}^{-},\chi )$. We
shift legs 1,2,5 as (\ref{shift1}) and 3,4 as (\ref{shift2}). The amplitude does not have any factorization poles, and the
only poles of $F_5(z)$ are soft poles. As the shifted amplitude vanishes for $\hat{p_{5}}
\rightarrow 0$ the only contributions come from $\hat{p_{1}}$ or $\hat{p_{2}}
\rightarrow 0$ soft limits. The residue at $z=1/a_1$ then reads 
\begin{equation}
\mathrm{Res}_{z=\frac{1}{a_{1}}}F_5(z)=- \frac{\widehat{A_{5}}|_{z= 1/a_{1}}}{(1-a_{2}/a_{1})(1-a_{5}/a_{1})}\,. \label{res}
\end{equation}
The value of the shifted amplitude $\widehat{A_{5}}|_{z=1/a_{1}}$
can be obtained from the soft theorem (\ref{softT}) by considering the
shifted kinematics, 
\begin{equation*}
\hat{p_{1}}=0,\quad \hat{p_{2}}=\Bigr( \frac{a_{1}-a_{2}}{a_{1}}\Bigl)
p_{2},\quad \hat{p_{5}}=\Bigr( \frac{a_{1}-a_{5}}{a_{1}}\Bigr) p_{5} .
\end{equation*}
Plugging the result into (\ref{res}) we get 
\begin{equation}
\mathrm{Res}_{z=\frac{1}{a_{1}}}F_5(z)=\frac{\mathrm{i}F_{0}}{F^{6}}\Bigl( F^{2}\,\mi\,\frac{F_{0}^{2}}{2}\Bigr) s_{25} .\label{t1}
\end{equation}
Similarly, the residue at the pole $z=\frac{1}{a_{2}}$ for $\hat{p_{2}}=0$ gives
\begin{equation}
\mathrm{Res}_{z=\frac{1}{a_{2}}}F_5(z)=\frac{\mathrm{i}F_{0}}{F^{6}}\Bigl( F^{2}\,\mi\,\frac{F_{0}^{2}}{2}\Bigr) s_{15},\label{t2}
\end{equation}
and after using the momentum conservation the sum of (\ref{t1}), (\ref{t2}) reproduces the
formula (\ref{amp5}).

\section{Uniqueness of the model}

\vspace{-0.2cm}

In the last part we turn the procedure around, and will reconstruct our
non-linear sigma model for $N=2$ as a unique theory which satisfies soft
theorem of the type (\ref{soft}). Following the logic of \cite{Cheung:2014dqa} 
we start with the ansatz for the amplitude of three types
of scalar fields $\phi^{\pm },\chi$ in terms of kinematical invariants and
impose the soft theorem of the general type 
\begin{equation}
\lim_{p_1\rightarrow 0}A_{n}=\sum_{i}c_{i}A_{n-1}^{(i)}  \label{soft3}
\end{equation}
as a constraint. If the right-hand side is zero we deal with the standard
Adler theorem (for more details see \cite{Cheung:2016drk}). 
To go beyond the standard situation we demand a
non-zero right-hand side when shifting charged particles, and keep the Adler
zero only for the neutral $\chi$. We went up to the 7pt amplitudes to check
that the unique answer is our model, $U(1)$-fibrated CP$(1)$, and the general $c_i$
constants are set in accordance with (\ref{soft2}).

The natural question is if there are more theories of this type for more
than three scalar fields beyond our explicit example (\ref{Lagr1}). This is
an open question, and we believe that this procedure is a very useful tool
to address the problem and potentially find new theories with non-trivial
soft theorems. In principle, we can also look at amplitudes for theories with only two
types of scalar fields. In the upcoming work \cite{future} we will prove that
for any such theory, under the assumption that the soft theorem (\ref{soft})
with $F_{I}^{J}=F\delta _{I}^{J}\neq 0$ is valid, and assuming non-vanishing
4pt amplitude, all the odd-particle amplitudes have to vanish. Therefore all
Goldstone boson amplitudes must necessarily have the Adler zero. This
supports the statement that the only non-linear sigma model for two scalars
are CP$(1)=O(3)/O(2)$ and $O(1,2)/O(2)$.

\vspace{-0.4cm}

\section{Conclusion}

\vspace{-0.2cm}

In this letter we found a new soft theorem for the Goldstone boson
amplitudes. Using the example of $SU(N)/SU(N-1)$ non-linear sigma models, we
showed that generically the amplitudes do not vanish in the soft limit but
rather reduce to a recursion. Explicit expressions are presented in the
simplest $N=2$ case which describes a pair of charged NGBs and a single
neutral NGB. We proved that this theory can be uniquely fixed from the
tree-level $S$-matrix if we impose the soft theorem as a constraint.
Consequently, we derived the recursion relations to reconstruct
all tree-level amplitudes.

Our work opens new avenues in studying NLSMs, and more generally EFTs using
non-vanishing soft limits of scattering amplitudes. In \cite{future} we will generalize this work, and use the soft theorems as the
theoretical tool to explore larger space of theories based on properties of their
scattering amplitudes. The exceptional EFTs also appear in the Cachazo-He-Yuan (CHY) formula \cite{CHY1}, ambitwistor strings \cite{CHY2} and the color-kinematics duality \cite{CHY3}, while the non-trivial soft limits have been encountered in the calculation of the leading non-zero term in the soft limit of $SU(N)$ NLSM amplitudes using the CHY formalism \cite{CHY4}. It would be fascinating to explore if our result fits into this framework.

\vspace{0.15cm}

\textit{Acknowledgment:} This work is supported in part by the Czech Government projects GACR
18-17224S and LTAUSA17069, by DOE grants No. DE-SC0009999 and  DE-SC0011842, and the funds
of University of California.

\end{document}